\documentclass[letterpaper]{article}
\usepackage{aaai}
\usepackage{times}
\usepackage{helvet}
\usepackage{courier}
\usepackage{graphicx,subcaption}
 \usepackage{url}

\begin{document}
%
\title{Towards Domain-Specific Characterization of Misinformation}
\author{Fariha Afsana\textsuperscript{1},
Muhammad Ashad Kabir\textsuperscript{1},
Naeemul Hassan\textsuperscript{2},
Manoranjan Paul\textsuperscript{1}\\
\textsuperscript{1}{School of Computing and Mathematics, Charles Sturt University, Australia}\\
\textsuperscript{2}{Philip Merrill College of Journalism, University of Maryland, USA}\\
fafsana@csu.edu.au,
akabir@csu.edu.au, 
nhassan@umd.edu, 
mpaul@csu.edu.au}
\maketitle
\maketitle
\begin{abstract}
\begin{quote}
The rapid dissemination of health misinformation poses an increasing risk to public health. To best understand the way of combating health misinformation, it is important to acknowledge how the fundamental characteristics of misinformation differ from domain to domain. 
This paper presents a pathway towards domain-specific characterization of misinformation so that we can address the concealed behavior of health misinformation compared to others and take proper initiative accordingly for combating it. With this aim, we have mentioned several possible approaches to identify discriminating features of medical misinformation from other types of misinformation. Thereafter, we briefly propose a research plan followed by possible challenges to meet up. The findings of the proposed research idea will provide new directions to the misinformation research community.  
\end{quote}
\end{abstract}

\section{Introduction}
\noindent Internet today is considered as an open door to the whole world of information. According to a Report from the International Telecommunication Union~\cite{IUT}, today 3.9 billion people are using the internet, representing approximately half of the world's population (7.7 billion).
Besides getting information from the internet, people can easily publish contents on blogs, social media or websites making it harder to differentiate between reality and fiction. Many content creators use social media sites or any other means of online platforms for their own advantages (e.g., intention to reach large audiences) and thus give rise to the spread of misinformation. Though all information over the Internet is not reliable, it is getting more and more believers at every instant. According to research~\cite{pew}, a majority of U.S. adults - 62\% - depending on news from social media which vary in the scale of legitimacy. However, it is clear that the way our access to the Internet is being mediated, we are becoming more vulnerable to a seemingly unstoppable force of misinformation.

The proliferation of misinformation dissemination can have devastating consequences to our wealth, democracy, health and national security~\cite{p1}. For example, during the 2016 US presidential election campaign, hundreds of falsified and heavily biased stories were spread over various websites which had influence over the poll~\cite{p18}~\cite{p19}. In 2016, the market value of battery manufacturer ‘Samsung SDI’ was dropped sharply by more than half a billion dollars after being tweeted by Tesla boss Elon Musk that the company was working with Panasonic on its next electric car which was a misinformation\footnote{\url{https://www.bbc.com/news/business-48871456}}. One of the key consequences of misinformation dissemination on health domain is vaccine hesitancy which arouses from the measles‐mumps‐rubella (MMR) - autism controversy by ``Andrew Wakefield’s study''~\cite{p20}. Chang et al.~\cite{p21} reported that this controversy resulted in a deterioration in the MMR immunization rates and negative spillover onto other vaccines in the US which is alarming. It seems that medical misinformation can have a detrimental impact on public health. 

 
While misinformation dissemination in politics requires great attention, it might demand greater attention when it is the case with medical advice. Political misinformation can be threatening to democracy, business misinformation may have a detrimental effect on the reputation and brand of an institution or company but medical misinformation can threat our lives to the death. In this internet age, people are dependent more on the internet search for seeking medical help rather than going to medical professionals. Research has found that 80\% of people self-diagnose them by the help of “Dr. Google”~\cite{p22}. As lies disseminate faster than the truth~\cite{p1004}, people get misguided easily by medical misinformation posing their physical health at risk. 
Consumption of health misinformation often scares people unnecessarily about their health and gradually builds  misconceptions about treatments which may result in a delay in adopting necessary medical care and attention on time. Besides, an individual may get motivated to spend money on treatments that are not accurate or medically proven. 
 For example, many cancer patients often turn to complementary medicine being convinced by internet-based misinformation~\cite{2}~\cite{3} which might hinder timely prevention and treatment decision. It can indulge the trustworthiness of the health-care providers and create a lack of trust in taking medicine, food, and vaccinations~\cite{m2}. As a whole, misinformation in the health sector not only poses risks to public health but also creates problems for the health care society. There may exist more than one factor (e.g., financial gain, political benefit, misconception) behind the motive of spreading health misinformation. Therefore, combating misinformation from political or business domain may contribute to combat health misinformation partially. Hence, our aim is to identify the similarities and dissimilarities of misinformation among different domains so that we can identify the main causes of health misinformation dissemination.

Various research has been proposed in the literature to combat misinformation most of which are conducted for political perspective~\cite{f1}~\cite{f2}~\cite{f3}. Recently, the medical domain has gained a lot of attention for being contaminated with misinformation~\cite{m1}~\cite{m2}~\cite{m3}~\cite{p1001}~\cite{p1002}. Ghenai et al.~\cite{m1} presented a case study of health misinformation on social media by looking into the characteristics of users propagating unverified cures of cancer on Twitter. In a recent study, Dhoju et al.~\cite{m2} has identified
structural, topical and semantic differences between health related news articles from reliable and unreliable media by conducting a systematic content analysis. Kinsora et al.~\cite{p1002} developed a new labelled dataset of misinformative and non-misinformative comments from a medical health forum, MedHelp, with a view to making a resource for medical research communities to study the spread of medical misinformation. Wang et al.~\cite{p1001} conducted a systematic review to identify the driving mechanism for health misinformation dissemination. They reviewed health issues identified, proposed frameworks and network models followed by empirical strategies. However, little attention has been paid to identify the root characteristics of health misinformation compared to other domains. 
Implementing machine learning to combat the dissemination of misinformation is admirable before which there is a need to identify the principal characteristics of misinformation to stop this harmful activity. Therefore, more research is needed to point out the root aspects of health misinformation so that it can be handled properly. Applying insights gained from previous works, results from our proposed idea will supplement present research by adding a new direction towards combating health misinformation.

To our view, a domain (e.g., medical, business, and politics) specific characterization of misinformation from machine learning perspective can help us in identifying the deviating features between health misinformation and other domains and thus conduct us to adopt appropriate countermeasures for combating health misinformation. 
From the above discussion, it can be stated that the motivation and impact of spreading misinformation differ from domain to domain. The patterns of dissemination, the motive of the spreader, perception of the user might be different from one domain to another. As a result, there might be structural, semantic or linguistic differences in misinformation among various domains. Our aim is to explore these differences by conducting extensive feature analysis. To accomplish this, we need to develop a properly annotated dataset containing misinformation from various domains (e.g., medical misinformation, political misinformation, etc.). This dataset will assist us to extract most discriminating features through analysis.
This paper provides our initial plan to investigate how misinformation in health domain differs from other domains such as political and business. 

\section{Research Plan}
It is theoretically apparent that the way medical misinformation disseminates to the user has a different impact with a different intention than that of political or business misinformation. To examine and analyze these differences in a logical and systematic manner, we are motivated to put forward the following research plan:

\subsection{Research Question (RQ)}
\textit{RQ: Are there any differences in the patterns of medical misinformation  compare to misinformation from other domains?}
\subsection{Hypothesis}
We have defined the hypotheses as follows:\\
\textbf{H1:} There are significant differences in the linguistic, semantic and structural patterns between medical misinformation and misinformation from other domains.\\
\textbf{H2:} There are differences in the behavior of users linked with the misinformation from different domains. \\
\textbf{H3:} The negative and positive persuasion of medical misinformation are different than those of other domains.\\
\textbf{H4:} The spreading pattern of medical misinformation differs from other domains.

\subsection{Research Challenge}
Obtaining a high-quality fine-grained dataset would be a great challenge for accomplishing our goal. Currently, there are many available datasets or fact-checking tools for measuring the level of political misinformation. But for healthcare applications, there is still a need for large medical information datasets from authentic sources. Again, identifying discriminative features of medical misinformation compared to misinformation from other domain will require both truthful and deceptive news in the corpus with a proper annotation. The scarcity of properly annotated deceptive news and truthful news for the medical information is a major stumbling block in this case.  
\subsection{Aims and Objectives}
To test the hypotheses our first objective is to formulate a fine-grained dataset with misinformation from different domains, their meta information, relation to the social media, etc. After having a well-structured dataset we will then devise all possible features relevant to each hypothesis and execute extensive analysis towards our vision.
\subsection{Research Design}
We have designed our research direction according to misinformation triangle. Misinformation requires three core factors - Motivation, Dissemination Services and Social Network- to be disseminated successfully and the absence of any of these factors  will make the spread more difficult. The collective representation of these three factors is known as misinformation triangle~\cite{p23}. 

Misinformation is always created with a specific purpose in mind. The answer to the question: `why' indicates the motivation behind misinformation spread. For manipulating and spreading misinformation, Social media networks are the most preferred way. But social media is not the only strategy being employed for misinformation spread. Numerous advertising support and services, automated bots, techniques for spammers are available to carry out the task of propagating misinformation. So, to conduct a successful analysis for combating misinformation, we must address the domain wise characteristics of misinformation based on these three core pillars. We have aligned following four approaches to be conducted to address our problem:
\begin{itemize}
    \item \textbf{Content Analysis} for understanding the structural resemblances and differences\\
        \item \textbf{Profiling User Behaviour} for understanding the motivational similarities and dissimilarities of spreader\\
    \item \textbf{Profiling domain-specific impact of misinformation} for understanding the persuasive differences of nature of parties or people linked to it\\
    \item \textbf{Profiling Network Behaviour} for understanding the similarities and differences in spreading pattern
\end{itemize}
\begin{figure}[hbt!]
 \centering
     \includegraphics[width=3.5in,height = 3.5in]{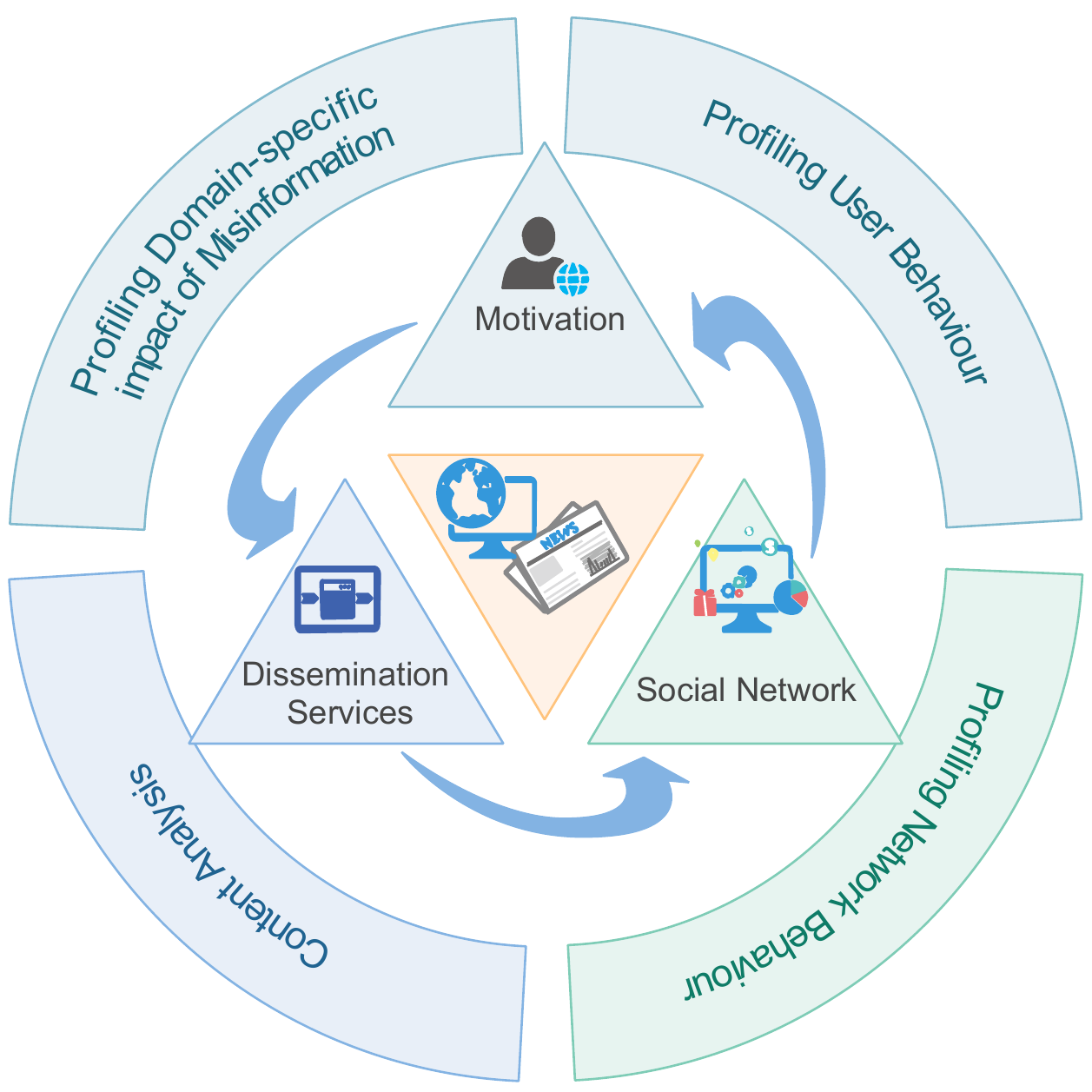}
    \caption{Possible approaches for domain-specific characterization based on misinformation triangle}
    \label{fig:f2}
\end{figure}

Fig. \ref{fig:f2} depicts the relationship among three factors of misinformation triangle and relevant direction of research for identifying differences among misinformation from various domains.

\section{Research Methodology}
In this section, we will describe possible pathways for implementing our visioned approaches. 

\subsection{Data Collection} 
As machine learning depends heavily on data, the first step towards accomplishing our goal would be data collection. Data acquisition and annotation are required for creating our dataset. We aim to create a dataset consisting of misinformation from different domains (e.g., medical misinformation, political misinformation, etc.) with a finely annotated label from authentic sources. Currently, this type of dataset is not available.
\subsection{Content Analysis}
According to our understanding, there might be structural differences among contents obtained from the medical domain and some other domains. So, given a well-structured dataset, we might conduct news content analysis including sentiment analysis, lexical and syntactic feature analysis, auxiliary information analysis associated with the text, etc. to extract discriminative characteristics of domain-specific misinformation.
\subsection{Profiling User Behaviour}
It is clear to us that the motivation behind misinformation spread varies from domain to domain. For example, intention for political misinformation can be swaying people for power while the intention of medical misinformation can be making the revenue for a business. In order to analyse user motivation behind misinformation dissemination, we can conduct user-based feature analysis including characteristics of user profile, individual feature, group feature, user demographics such as number of followers/followees in social media, age, emotion, user’s reaction towards misinformation such as supporting or denying, sensational reactions and sceptical opinions. 
\subsection{Profiling domain-specific impact of misinformation}
In order to gain a better insight of motivation of misinformation we will conduct an analysis on the persuasive behavior of parties involved with misinformation. A misinformation can influence people negatively or positively based on their nature of persuasion. Analyzing their prior knowledge, persuasive nature, belief, etc., can help us to measure the influence of misinformation.

\subsection{Profiling Network Behaviour}
Analyzing network properties and behavior can be a way to complement the other three approaches in identifying characteristics of misinformation. In order to identify the spreading pattern of misinformation, we will consider the trajectory of the spread of news, temporal dissemination of original and shared news, speed of news propagation, etc., as network-based features.

As a whole, we will identify the similarities and differences of characteristics of health misinformation compare to misinformation in other domains. This will provide a new research direction to explore combating health misinformation.
\section{Conclusion}

\noindent In this paper, we have presented a research direction for analyzing the characteristics of misinformation from different domains based on the conceptual illustration. We have aligned our proposed solutions with misinformation triangle to test our hypotheses. We have outlined some research directions and challenges for this vision. The need to realize this vision is seemingly evident. The findings from this research direction will help to combat health misinformation successfully.

\bibliography{main}

\begin{thebibliography}{}

\bibitem[\protect\citeauthoryear{Allcott and Gentzkow}{2017}]{p18}
Allcott, H., and Gentzkow, M.
\newblock 2017.
\newblock {Social Media and Fake News in the 2016 Election}.
\newblock {\em Journal of Economic Perspectives} 31(2):211--236.

\bibitem[\protect\citeauthoryear{Carrieri, Madio, and Principe}{2019}]{p20}
Carrieri, V.; Madio, L.; and Principe, F.
\newblock 2019.
\newblock {Vaccine Hesitancy and Fake News: Quasi-experimental Evidence from
  Italy}.
\newblock Health, Econometrics and Data Group (HEDG) Working Papers 19/03,
  HEDG, c/o Department of Economics, University of York.

\bibitem[\protect\citeauthoryear{Chang}{2018}]{p21}
Chang, L.~V.
\newblock 2018.
\newblock {Information, education, and health behaviors: Evidence from the MMR
  vaccine autism controversy}.
\newblock {\em Health Economics} 27(7):1043--1062.

\bibitem[\protect\citeauthoryear{Dhoju \bgroup et al\mbox.\egroup }{2019}]{m2}
Dhoju, S.; Rony, M. M.~U.; Kabir, M.~A.; and Hassan, N.
\newblock 2019.
\newblock Differences in health news from reliable and unreliable media.
\newblock In {\em Companion Proceedings of The 2019 World Wide Web Conference
  on - WWW '19},  981--987.
\newblock New York, New York, USA: ACM Press.

\bibitem[\protect\citeauthoryear{Downer \bgroup et al\mbox.\egroup }{1994}]{2}
Downer, S.~M.; Cody, M.~M.; McCluskey, P.; Wilson, P.~D.; Arnott, S.~J.;
  Lister, T.~A.; and Slevin, M.~L.
\newblock 1994.
\newblock {Pursuit and practice of complementary therapies by cancer patients
  receiving conventional treatment}.
\newblock {\em BMJ} 309(6947):86--89.

\bibitem[\protect\citeauthoryear{Ghenai and Mejova}{2018}]{m1}
Ghenai, A., and Mejova, Y.
\newblock 2018.
\newblock {Fake Cures: User-centric Modeling of Health Misinformation in Social
  Media}.
\newblock {\em Proceedings of the ACM on Human-Computer Interaction} 2:1--20.

\bibitem[\protect\citeauthoryear{Gottfried and Shearer}{2016}]{pew}
Gottfried, J., and Shearer, E.
\newblock 2016.
\newblock News use across social media platforms 2016.
\newblock Technical report, Pew Research Center.

\bibitem[\protect\citeauthoryear{Gu, Kropotov, and Yarochkin}{2017}]{p23}
Gu, L.; Kropotov, V.; and Yarochkin, F.
\newblock 2017.
\newblock The fake news machine: How propagandists abuse the internet and
  manipulate the public.
\newblock Technical report, Cyentia Institute, Cybersecurity Research and
  Publications Library.

\bibitem[\protect\citeauthoryear{Hassan \bgroup et al\mbox.\egroup }{2017}]{p1}
Hassan, N.; Arslan, F.; Li, C.; and Tremayne, M.
\newblock 2017.
\newblock {Toward Automated Fact-Checking}.
\newblock In {\em Proceedings of the 23rd ACM SIGKDD International Conference
  on Knowledge Discovery and Data Mining - KDD '17},  1803--1812.
\newblock New York, New York, USA: ACM Press.

\bibitem[\protect\citeauthoryear{{ICT Data} and {Statistics Division of
  IUT}}{2018}]{IUT}
{ICT Data}, and {Statistics Division of IUT}.
\newblock 2018.
\newblock Measuring the information society report 2018.
\newblock Technical report, International Telecommunication Union.
\newblock (Accessed: 2019-07-21).

\bibitem[\protect\citeauthoryear{Kinsora \bgroup et al\mbox.\egroup
  }{2017}]{p1002}
Kinsora, A.; Barron, K.; Mei, Q.; and Vydiswaran, V.~V.
\newblock 2017.
\newblock Creating a labeled dataset for medical misinformation in health
  forums.
\newblock In {\em 2017 IEEE International Conference on Healthcare Informatics
  (ICHI)},  456--461.

\bibitem[\protect\citeauthoryear{K{\l}ak \bgroup et al\mbox.\egroup
  }{2017}]{p22}
K{\l}ak, A.; Gawi{\'{n}}ska, E.; Samoli{\'{n}}ski, B.; and Raciborski, F.
\newblock 2017.
\newblock {Dr Google as the source of health information – the results of
  pilot qualitative study}.
\newblock {\em Polish Annals of Medicine} 24(2):188--193.

\bibitem[\protect\citeauthoryear{Langin}{2018}]{p1004}
Langin, K.
\newblock 2018.
\newblock {Fake news spreads faster than true news on Twitter—thanks to
  people, not bots}.
\newblock
  \url{http://www.sciencemag.org/news/2018/03/fake-news-spreads-faster-true-news/-twitter-thanks-people-not-bots}.
\newblock Accessed: 2019-07-01.

\bibitem[\protect\citeauthoryear{Lazer \bgroup et al\mbox.\egroup }{2018}]{f2}
Lazer, D. M.~J.; Baum, M.~A.; Benkler, Y.; Berinsky, A.~J.; Greenhill, K.~M.;
  Menczer, F.; Metzger, M.~J.; Nyhan, B.; Pennycook, G.; Rothschild, D.;
  Schudson, M.; Sloman, S.~A.; Sunstein, C.~R.; Thorson, E.~A.; Watts, D.~J.;
  and Zittrain, J.~L.
\newblock 2018.
\newblock {The science of fake news}.
\newblock {\em Science} 359:1094--1096.

\bibitem[\protect\citeauthoryear{Matthews \bgroup et al\mbox.\egroup
  }{2003}]{3}
Matthews, S.~C.; Camacho, A.; Mills, P.~J.; and Dimsdale, J.~E.
\newblock 2003.
\newblock {The Internet for Medical Information About Cancer: Help or
  Hindrance?}
\newblock {\em Psychosomatics} 44(2):100--103.

\bibitem[\protect\citeauthoryear{Murphy}{2019}]{p19}
Murphy, J.
\newblock 2019.
\newblock Evaluating information: Fake news in the 2016 us elections.
\newblock \url{https://libraryguides.vu.edu.au/c.php?g=460840{\&}p=5330649}.
\newblock Accessed: 2019-07-01.

\bibitem[\protect\citeauthoryear{Samuel and Za{\"{i}}ane}{2018}]{m3}
Samuel, H., and Za{\"{i}}ane, O.
\newblock 2018.
\newblock {MedFact: Towards Improving Veracity of Medical Information in Social
  Media Using Applied Machine Learning}.
\newblock Springer, Cham.
\newblock  108--120.

\bibitem[\protect\citeauthoryear{Sharma \bgroup et al\mbox.\egroup }{2019}]{f3}
Sharma, K.; Qian, F.; Jiang, H.; Ruchansky, N.; Zhang, M.; and Liu, Y.
\newblock 2019.
\newblock Combating fake news: A survey on identification and mitigation
  techniques.
\newblock {\em ACM TIST} 10:21:1--21:42.

\bibitem[\protect\citeauthoryear{Shu \bgroup et al\mbox.\egroup }{2017}]{f1}
Shu, K.; Sliva, A.; Wang, S.; Tang, J.; and Liu, H.
\newblock 2017.
\newblock {Fake News Detection on Social Media}.
\newblock {\em ACM SIGKDD Explorations Newsletter} 19(1):22--36.

\bibitem[\protect\citeauthoryear{Wang}{2018}]{p1001}
Wang, Y.
\newblock 2018.
\newblock {Systematic Review on the Social Mechanism of Health Misinformation
  Dissemination in the Internet Era}.
\newblock {\em European Journal of Public Health} 28.

\end{thebibliography}
\bibliographystyle{aaai}

\end{document}